\documentclass{article}

\usepackage[T1]{fontenc}
\usepackage[utf8]{inputenc}
\usepackage{amsmath}
\usepackage{amssymb}
\usepackage[toc,page]{appendix}
\usepackage{graphicx} 
\usepackage{dcolumn}  
\usepackage{bm}       
\usepackage{float}
\usepackage{authblk}

\newcommand{\be}{\begin{equation}}
\newcommand{\ee}{\end{equation}}
\newcommand{\bee}{\begin{equation*}}
\newcommand{\eee}{\end{equation*}}
\newcommand{\ud}{\mathrm{d}}
\newcommand{\ic}{\mathrm{i}}

\newcommand{\abs}[1]{\left\lvert#1\right\rvert}

\title{Localized solutions of nonlinear network wave equations} 

\author[1]{\normalsize{J. G. Caputo }\thanks{caputo@insa-rouen.fr}}
\author[1]{I. Khames \thanks{imene.khames@insa-rouen.fr}}
\author[1]{A. Knippel \thanks{arnaud.knippel@insa-rouen.fr}}

\affil[1]{Laboratoire de Math\'ematiques, INSA de Rouen Normandie\\ 76801 Saint-Etienne du Rouvray, France.}

\author[2]{\normalsize{A. B. Aceves} \thanks{aaceves@mail.smu.edu}}
\affil[2]{Department of Mathematics, Southern Methodist University, Dallas, TX 75275, USA.}

\date{\ }

\begin{document}
\maketitle
\vspace{-1cm}

\begin{abstract}
We study localized solutions for the nonlinear graph wave equation on finite arbitrary networks. 
Assuming a large amplitude localized initial condition on one node of the graph,
we approximate its evolution by the Duffing equation. 
The rest of the network satisfies a linear system forced by the excited node.
This approximation is validated
by reducing the nonlinear graph wave equation 
to the discrete nonlinear Schr\"odinger equation and by Fourier analysis.
Finally, we examine numerically the condition for localization in the parameter plane, 
coupling versus amplitude and show that the localization amplitude depends on the maximal
normal eigenfrequency.
\end{abstract}

\maketitle

\section{Introduction}

Intrinsic localized modes, also called discrete breathers are time periodic 
and (typically exponentially) spatially localized solutions that 
appear in nonlinear discrete systems arising in many physical, biological 
systems and networks. They were analytically studied first by Sievers 
and Takeno \cite{st88}, Page \cite{page}. 
Later, MacKay and Aubry \cite{MacKay94} proved the existence of discrete 
breathers 
by considering a lattice model of coupled anharmonic oscillators in 
the limit of very weak interaction (anticontinuous limit).
Localized solutions exist in nonlinear networks because of
the interplay between nonlinearity and discreteness. 
In fact, the non-resonance of the breather frequency with 
the linear spectrum is a 
necessary condition for obtaining a time-periodic localized state 
\cite{flach94}.

Localized modes have been investigated theoretically and numerically 
for a variety of physical systems \cite{flach}. 
Experimentally observed reports for various systems include 
Josephson-junction arrays \cite{Binder}, 
optical waveguides \cite {aceves96,eisenberg98}, 
photonic crystals \cite{fleischer03},
DNA double strand \cite{peyrard},
micromechanical oscillators \cite{sievers03} and
electronic circuits \cite{sievers07}. 
Recently, photonic lattices built with different configurations such as 
multi-core fibers \cite{christodoulides16,aceves16} and waveguide arrays \cite{eisenberg98,christodoulides16}, 
allow the enhancement of cubic (Kerr), quadratic, photorefractive, 
local and non-local nonlinearities \cite{weinstein17} 
under different discrete geometries \cite{vicencio13}. 
Novel graphene arrays \cite{kou16} suggest the existence of localized 
modes in the nano-meter scale. In all cases the underlying periodicity 
or discreteness leads to new families of optical discrete modes different
from the ones of continuous systems. 
Localized modes have been thoroughly studied in the Fermi-Pasta-Ulam lattice 
\cite{flach05} 
and in the discrete nonlinear Schr\"odinger equation \cite{panos10}.
They have been found in the dynamics of such complex objects as molecules
\cite{ovchinnikov}.

Here we aim to describe the dynamics
of arbitrary topology networks. The physical systems are
electrical networks of nonlinear elements, for example diodes or
Josephson junctions connected by inductances and systems of identical
particles coupled by mechanical interactions (see the book by Scott
\cite{Scott1} for examples) and fluid flows on a network \cite{maas}. 
The general model for these systems is the graph wave equation, 
where the usual continuum Laplacian is replaced by the graph Laplacian 
\cite{cvetkovic}. It arises naturally from discrete conservation 
laws \cite{cks13}. The graph Laplacian, being a symmetric positive matrix, has
real eigenvalues and we can choose a basis of orthogonal eigenvectors
corresponding to the normal modes \cite{cks13} giving rise to 
periodic solutions.
When nonlinearity is present, these normal modes generally couple together.
Only special eigenvectors with coordinates in $\{1,-1,0\}$ extend into
nonlinear periodic orbits \cite{ckkp17}. Other nonlinear solutions
exist for lattices, in particular the ones that are localized.
It is then natural to study them in general networks.

In the present article, we study the nonlinear graph wave
equation on a network of arbitrary topology
and search for nonlinear localized solutions by introducing
a large initial condition on only one of the nodes of network. 
We observe localized modes, 
i.e. large amplitude nonlinear excitations that do not decay significantly
over thousands of periods. The Fourier spectrum reveals the
presence of linear modes that are not eigenfrequencies of the
Laplacian but of a reduced matrix that we characterize. Our main
result is a full picture of a discrete breather on a general network.
The excited node satisfies a Duffing equation while the rest of
the network obeys a driven linear system which we present.
We confirm this approximation by numerical simulations and 
by modulation theory
for a natural frequency $\omega\neq 0$. Then the 
graph nonlinear wave equation reduces to a graph 
nonlinear Schr\"odinger equation. 
We illustrate the analysis by considering
two graphs, the paw graph and cycle 6. 
The localization threshold was estimated in the parameter plane:
coupling versus amplitude. As expected, waves delocalize as the
coupling increases because we get closer to the resonance condition.
However, different
nodes behave differently depending on their degree and the graph 
configuration.  \\
The article is organized as follows: 
We introduce the graph nonlinear wave equation and 
the localized modes in section \ref{sec2}. 
In section \ref{sec3},  we reduce the graph nonlinear wave equation for $\omega\neq 0$ 
to a discrete nonlinear Schr\"odinger equation and determine nonlinear 
localized solutions. Section \ref{sec4} confirms this analysis by studying
the dynamics in real and Fourier space of
two main networks; in particular we examine the localization vs delocalization
regimes in the parameter plane coupling vs amplitude. 
Section \ref{sec5} summarizes the paper.

\section{The graph nonlinear wave equation : Localized modes}
\label{sec2}

We study the nonlinear wave equation on a 
connected graph $\mathcal{G}$ with $N$ nodes
\begin{equation} \label{phi4}
{\ud ^2 \mathbf{u} \over \ud t^2}=\left(-\omega^2 \mathbf{I} 
+\epsilon \mathbf{\Delta}\right) \mathbf{u} - \mathbf{u}^3 ,
\end{equation}
where $\mathbf{u}=(u_1(t),u_2(t),\dots,u_N(t))^{T}$ is the field amplitude, 
$\mathbf{u}^3=(u_1^3,u_2^3,\dots,u_N^3)^{T}$,
$\mathbf{I}$ is the identity matrix and
$\mathbf{\Delta}$ is the graph Laplacian \cite{cvetkovic}. This $N\times N ~ $
matrix is $\mathbf{\Delta}=\mathbf{A}-\mathbf{D}$,
where $\mathbf{A}$ is the adjacency matrix such that $A_{ij}=1$ 
if nodes $i$ and $j$ are connected ($i\neq j$) and $A_{ij}=0$ otherwise,
and $\mathbf{D}$ is the diagonal matrix where the entry 
$d_{i}=\sum_{j=1}^{N} A_{ij}$ is the degree of vertex $i$.
For weighted graphs the $1$ in the adjacency matrix will be replaced
by the weight $\alpha_{ij}$ of the edge $\it{e}_{i,j}$.
The other parameters are the natural frequency $\omega$
and the linear coupling coefficient $\epsilon$ which is assumed 
small in this article. 
{Equation (\ref{phi4}) is well-posed because the cubic on-site 
nonlinearity guarantees the 
existence of the solution for all times.}
Note that we use bold-face capitals for matrices and bold-face lower-case letters for vectors.

Equation (\ref{phi4}) is an extension to a graph of the $\Phi^4$
well-known model in condensed matter physics \cite{Scott1}.
In the literature, for example \cite{Scott2} and references therein, 
the discrete $\Phi^4$ model was studied only in the particular case of lattices
where the graph Laplacian $\left(\mathbf{\Delta} \mathbf{u} \right)_i=u_{i+1}-2u_i+ u_{i-1}$
(for a one-dimensional lattice)
is a finite difference discretization of the continuous Laplacian.
This formulation is natural since the linear graph wave equation 
arises from discrete conservation laws \cite{cks13}.
We formulate the discrete $\Phi^4$ model using the graph Laplacian
to describe general networks of arbitrary topology, like for example
an electrical network.
The model can describe networks of nonlinear oscillators, such as
Josephson junctions or diodes \cite{Scott2}. In mechanical 
engineering, several aerospace structures e.g. turbine rotors 
or space antennas,
are composed of weakly coupled sectors assembled in a cyclic and 
symmetric configuration. Such a complex system can be 
reduced to (\ref{phi4}) \cite{grolet}.

The graph Laplacian $\mathbf{\Delta}$ is a real symmetric 
negative-semi definite matrix. It has real non positive 
eigenvalues $-\nu_i^2$ where 
\bee
\nu_1^2 =0 < \nu_2^2 \leq \dots \leq \nu_N^2  .
\eee
In our previous work \cite{ckkp17},
we constructed nonlinear periodic orbits which are extension
of some linear normal modes (associated to the eigenvectors) 
of the graph Laplacian.
Here instead, we take a different approach, we
assume a large amplitude localized initial condition
and search for nonlinear localized solutions.
An important remark is that 
this work can be generalized to any odd
power of the nonlinearity.

\subsection{Natural frequency $\omega=0$}

We consider a large amplitude initial condition localized  at node
$j$ and examine its evolution. 
First, we consider the anti-continuum limit,
$\epsilon = 0$, the evolution of $u_j$ 
satisfies
\begin{equation}
\label{uj0}
{\ud^2 u_j \over \ud t^2}=-u_j^3 ,
\end{equation}
where $u_j(0) = \rho $. The other nodes $u_k$, verify
$u_k(0)=0$ and therefore $u_k(t)=0$. 
The solution of (\ref{uj0}) can be written in terms of the Jacobi 
cosine elliptic function \cite{Abramowitz} 
\begin{equation}
\label{uj0cn}
u_j(t)=\rho ~ \mathrm{cn}\left(\rho t,{1\over \sqrt{2}}\right),
\end{equation}
where the modulus of $\mathrm{cn}$ is $\kappa= {1\over \sqrt{2}}$  
(Appendix \ref{jacobi}) and 
where we assumed ${\ud u_j \over \ud t}(0)=0$.
The period of oscillations (Appendix \ref{appendix_period}) is
\be\label{T0} 
T_{0} = {\Gamma^2\left({1\over 4}\right) \over \rho \sqrt{\pi}}, \ee
where $\Gamma(.)$ is the gamma function and 
$\Gamma\left({1\over 4}\right) \approx 3.6256$. 
The frequency of oscillations is
\be\label{Omega0}
\Omega_{0}={2\pi \over T_{0}}={2\pi\sqrt{\pi}  \over \Gamma^2\left({1\over 4}\right)} \rho .
\ee

Now examine the weak coupling limit $\epsilon \ll 1$. 
The nearest neighbors $k$ of $j$ solve the forced system 
\begin{equation}
 {\ud^2 u_{k} \over \ud t^2}= \epsilon \sum_{p=1}^{N} \Delta_{kp} u_p  -u_{k}^3 
 = -\epsilon d_{k} u_{k} +\epsilon u_j + 
\epsilon \sum_{p \sim k, ~ p\neq j} u_p-u_{k}^3,
\end{equation}
where $d_{k}$ is the degree of the node $k$,
the notation $p \sim k$ indicates the adjacency of vertices 
and the sum is taken over the other neighbors $p$ of $k$.
We assume that $u_{k}$ is small and will find a condition
on $\rho$ for this to hold.
If $u_{k}$ is small, it is natural to neglect the cubic term
$u_{k}^3$. The part of the solution for $u_k$ due to the
forcing is
\be {\ud^2 u_k^f \over \ud t^2}=\epsilon u_j(t) = \epsilon
\rho ~ \mathrm{cn}\left(\rho t,{1\over \sqrt{2}}\right) ,\ee
where the forcing $u_j$ is periodic of frequency $\Omega_0$ 
and amplitude $\rho$.
Then, the response $u_k$ to this periodic forcing 
will be of amplitude 
\be \label{amp_uk}
\abs{u_k^f} = \mathcal{O}\left(\frac{ \epsilon} {\rho} \right).\ee
This amplitude is small if $\rho \ge 1$. 
Similarly, the next nearest neighbors $l$ of node $j$ 
exhibit a forced oscillation given by 
${\ud^2 u_l^f \over \ud t^2}=\epsilon u_k(t)$ and this gives
\be \label{amp_ul}
\abs{u_l^f} = \mathcal{O} \left({\epsilon^2 \over \rho^3}\right).\ee

For simplicity and without loss of generality, we assume
an initial excitation of node $j=1$.
The evolution of the nodes $\{2,\dots,N\}$ is described by the forced system of linear ordinary differential equations
\begin{equation}
\label{un}
{\ud^2 \mathbf{v} \over \ud t^2} 
=\epsilon ~\mathbf{\Delta}^{1} ~\mathbf{v} + \mathbf{f},
\end{equation}
where
$\mathbf{v}=\left(u_{2},u_{3},\dots,u_{N} \right)^T$, $\mathbf{\Delta}^{1}$ is the matrix obtained by removing the first line and the first column from the graph Laplacian $\mathbf{\Delta}$
and where $\mathbf{f}=\left(f_1,f_2,\dots,f_{N-1} \right)^T$ is the forcing term such that 
$f_{k}=\epsilon u_1$ if $k$ adjacent to $1$ ($k\sim 1$)
and $0$ otherwise. 
The matrix $\mathbf{\Delta}^{1}$ is a reduction of the
graph Laplacian $\mathbf{\Delta}$. It is therefore real symmetric and negative, 
then $\mathbf{\Delta}^{1}$ has real eigenvalues $0>-\omega_1^2 \geq-\omega_2^2 \geq \cdots \geq -\omega_{N-1}^2$ and a basis of orthonormal eigenvectors $\mathbf{z}^{1},\mathbf{z}^{2}, \dots \mathbf{z}^{N-1}$. These verify
\bee
\mathbf{\Delta}^{1} ~\mathbf{z}^{m}=-\omega_{m}^{2} ~\mathbf{z}^{m},
\eee
for $m \in \{1,\dots,N-1 \}$.
We expand $\mathbf{v}$ using a basis of the eigenvectors 
$\mathbf{z}^{m}$ as 
\be
\label{v}
\mathbf{v} = \sum_{m=1}^{N-1} a_m ~\mathbf{z}^{m}.\ee
Substituting (\ref{v}) into (\ref{un}) and projecting on each eigenvector $\mathbf{z}^{m}$, 
we get 
\bee
{\ud^2 a_m \over \ud t^2}= -\epsilon \omega_m^2 a_m +\sum_{p=1}^{N-1} f_p z^m_{p},
\eee
where we have used the orthonormality of the eigenvectors of $\mathbf{\Delta}^{1}$.
The sum can be written as 
$$\sum_{p=1}^{N-1} f_p z^m_{p} = \epsilon u_1 \sum_{k \sim 1}  z^m_{k-1}.$$
We then get a set of $(N-1)$ second order inhomogeneous ordinary 
differential equations
\be
\label{alpha}
{\ud^2 a_m \over \ud t^2}= -\epsilon \omega_m^2 a_m +\epsilon u_1 \sum_{k \sim 1}  z^m_{k-1},
\ee
where $m \in \{1,\dots,N-1 \}$. 
At this level, we just rewrote equation (\ref{un}) in the basis
$\mathbf{z}^{m}$. Initially, $a_m(0)={\ud a_m \over \ud t}(0)=0$ so that
we only observe the forced response of the system. In particular,
the modes $a_m$ such that $\sum_{k \sim 1}  z^m_{k-1}=0$ 
will remain zero. 
This reveals that the harmonic frequencies of the solutions 
$u_2,\dots,u_N$ that will be observed are
\be
\label{freqnn_w0}
\sqrt{\epsilon}~ \omega_m, 
\ee
for $m\in \{1,\dots,N-1 \}$ such that $\sum_{k \sim 1}  z^m_{k-1} \neq 0$.
{This is the sum of the eigenvector components on the neighboring nodes of the excited node.}

In general, for initial excitation of node $j$, the harmonic frequencies (\ref{freqnn_w0})
of the matrix $\mathbf{\Delta}^j$ (obtained by removing the $j$ line and the $j$ column from the graph Laplacian $\mathbf{\Delta}$)
will be observed if
\be \label{cond_freq}
\sum_{k\sim j,~ k<j} z_k^m +\sum_{k\sim j,~ k>j} z_{k-1}^m \neq 0, 
\ee
where $\mathbf{z}^m$ are the eigenvectors of $\mathbf{\Delta}^j$.
This condition on the eigenvectors of $\mathbf{\Delta}^j$ depends 
on the topology. For example, it could be not satisfied when 
there are symmetries 
in the graph.  Adding weights will usually break the symmetries, 
as shown below. Then the condition would be satisfied for all 
eigenvectors and all eigenfrequencies would be observed.

Following the interlacing theorem \cite{Fisk},
the eigenvalues 
$-\omega_1^2,-\omega_2^2,\dots,-\omega_{N-1}^2$ of $\mathbf{\Delta}^j$,
the submatrix of the graph Laplacian $\mathbf{\Delta}$,
interlace the eigenvalues of $\mathbf{\Delta}$,
$-\nu_1^2,-\nu_2^2,\dots,-\nu_N^2$ as
\be
\nu_1^2 \leq \omega_1^2 \leq \nu_2^2 \leq \omega_2^2 \leq \nu_3^2 \leq \dots \leq \nu_{N-1}^2 \leq \omega_{N-1}^2 \leq \nu_N^2.
\ee
For a general graph, the spectrum of the graph Laplacian $\mathbf{\Delta}$ needs to be computed numerically and 
the spectrum of $\mathbf{\Delta}^j$ as well. 
For some special cases however, like cycles, chains and grids, the 
eigenvalues and 
eigenvectors of the Laplacian have an explicit formula. 
In \cite{ckk18}, 
we determined a part of the spectrum of $\mathbf{\Delta}$ for 
special configurations, through the graph topology.

\subsection{Natural frequency $\omega\neq 0$}
Now, we consider the equation (\ref{phi4}) with a natural frequency $\omega \neq 0$. 
In the anti-continuum limit $\epsilon=0$, the evolution at the excited node $j$ satisfies
\begin{equation}
{\ud^2 u_j \over \ud t^2}=-\omega^2 u_j -u_j^3 ,
\end{equation}
where $u_j(0)=\rho$.
The solution can be written in terms of cosine elliptic functions \cite{Abramowitz}
\begin{equation}
u_j(t)=\rho ~ \mathrm{cn}\left(\sqrt{\omega^2 + \rho^2}~ t,~\kappa \right),
\end{equation}
where the modulus $\kappa=\sqrt{\rho^2 \over 2\left(\omega^2 +\rho^2\right) }$ and we assumed ${\ud u_j \over \ud t}(0)=0$.\\
As above, we examine the weak coupling limit $\epsilon \ll 1$ and assume an excitation at node $j=1$. 
Following the same procedure, the evolution at nodes $\{2,\dots,N\}$
is described by the forced system of linear ordinary differential equations
\begin{equation}
\label{unw}
{\ud^2 \mathbf{v} \over \ud t^2} 
=\left( \epsilon ~\mathbf{\Delta}^{1} -\omega^2 ~\mathbf{I}\right) \mathbf{v} + \mathbf{f}.
\end{equation}
Substituting (\ref{v}) into (\ref{unw}) and projecting on each eigenvector $\mathbf{z}^{m}$ of $\mathbf{\Delta}^{1}$, 
we get 
\be
{\ud^2 a_m \over \ud t^2}= -\left(\epsilon \omega_m^2 + \omega^2 \right) a_m +\epsilon u_1 \sum_{k \sim 1}  z^m_{k-1}, 
\ee
where $m\in \{1,\dots,N-1 \}$.
The harmonic frequencies are
\be
\label{freqnn_phi4}
\sqrt{\epsilon \omega_m ^2 + \omega ^2 }, 
\ee
for $m\in \{1,\dots,N-1 \}$ such that $\sum_{k \sim 1}  z^m_{k-1} \neq 0$.

\section{Modulation theory $\omega \ne 0$}
\label{sec3}

When the natural frequency is not zero, following \cite{peyrard92},
we reduce the discrete $\Phi^4$ equation (\ref{phi4}) to the discrete nonlinear Schr\"odinger equation.
We write 
\be\label{psi}
\mathbf{u}(t) = \sqrt{\epsilon}\boldsymbol{\psi} (T) \rm{e}^{i \omega t} 
+ \sqrt{\epsilon}\boldsymbol{\psi}^{*} (T) \rm{e}^{-\ic \omega t} ~, 
\ee
where $T=\epsilon t$, 
$\boldsymbol{\psi}=(\psi_1(t),\psi_2(t),\dots,\psi_N(t))^{T}$ is the field vector and 
$\boldsymbol{\psi}^{*}$ is the complex conjugate of $\boldsymbol{\psi}$.
Plugging (\ref{psi}) into (\ref{phi4}) and collecting terms in order
of $\epsilon ^{1\over 2}, \epsilon^{3 \over 2}, \dots$, we obtain for the
order $\epsilon^{3 \over 2}$ the graph nonlinear Schr\"odinger equation 
(see Appendix \ref{NLS})
\be\label{gnls}
{2 \ic \omega \over \epsilon} {\ud \boldsymbol{\psi} \over \ud t}= 
\mathbf{\Delta} \boldsymbol{\psi} 
- 3 \abs{\boldsymbol{\psi}}^2 \boldsymbol{\psi} ~.
\ee
This model describes the coupling 
between waveguides in an optical array. 
In \cite{ac14}, we examined how linear normal modes couple due to the cubic 
nonlinearity in (\ref{gnls}). Here instead, we assume 
a large amplitude localized initial condition.
This is a natural and relevant consideration that 
parallels classical studies of 
discrete solitons in the nonlinear Schr\"odinger equation
and light localization in nonlinear photonic structures.

We assume that $\abs{\psi_j}= r={\rho \over 2 \sqrt{\epsilon}} \geq 1$ 
constant at a given node where $\rho=u_j(0)$, and $\abs{\psi_k}=0,~\forall k\neq j$.
The evolution of the excited node $j$ is given by
\begin{equation}
\label{psij}
{2 \ic \omega \over \epsilon }{\ud \psi_j \over \ud t}= 
- 3 r^2 \psi_j .
\end{equation}
The solution of (\ref{psij}) is
\begin{equation*}
\psi_j (t) = r~ \mathrm{e}^{\ic {3 \epsilon r^2\over 2 \omega } t}
= {\rho \over 2 \sqrt{\epsilon}} \mathrm{e}^{{\ic 3 \rho^2 \over 8 \omega}  t}.
\end{equation*}
Thus, the solution $u_j$ can be approximated using (\ref{psi}) by
\begin{equation*}
u_j (t)\approx  \rho ~\cos \left( \left({3 \rho^2 \over 8 \omega}+\omega \right) t \right),
\end{equation*}
where the nonlinear frequency is
\begin{equation}
\label{Omega}
\Omega \approx {3 \rho^2 \over 8 \omega}+\omega .
\end{equation}
This regime is valid when the correction to the frequency
of oscillation due to the nonlinearity is smaller than the natural frequency
\be \label{cond_gnls} {3 \over 8 \omega} \rho^2 \ll \omega . \ee
This means $\omega$ large enough.
{Discrete breathers for (\ref{gnls}) in chains were studied by
Panayotaros in \cite{panos10}.
He used a continuation argument in $\epsilon$ starting from the 
anticontinuous limit,
to show the existence of discrete breathers.}

The nearest neighbors $k$ of $j$ solve a forced system 
\begin{equation}
{2 \ic \omega \over \epsilon }{\ud \psi_{k} \over \ud t} = 
\sum_{p=1}^{N} \Delta_{kp} \psi_p 
=-d_{k} \psi_{k} + \psi_j +\sum_{p \sim k, ~ p\neq j} \psi_p,
\end{equation}
where we neglected the cubic terms. A particular solution of the forced part 
${2 \ic \omega \over \epsilon }{\ud \psi_{k}^f \over \ud t}= \psi_j$ is
\be
\psi_{k}^f(t)={1\over 3 r}\left(1-\mathrm{e}^{\ic {3 \epsilon r^2 \over 2 \omega} t} \right)
=
{2 \sqrt{\epsilon}\over 3 \rho}
\left(1- \mathrm{e}^{\ic {3 \rho^2 \over 8 \omega} t}\right).
\ee
Similarly, the evolution of the next nearest neighbors $l$ is given by
\begin{equation}
{2 \ic \omega \over \epsilon }{\ud \psi_{l} \over \ud t}= 
\sum_{p=1}^{N} \Delta_{lp} \psi_p =
-d_{l} \psi_{l} + \psi_{k} +\sum_{p \sim l,  ~ p\neq k} \psi_p.
\end{equation} 
We have the proportional scalings for $\psi_k$ and $\psi_l$
\bee
\abs{\psi_{k}} =\mathcal{O} \left( {\sqrt{\epsilon}\over \rho} \right)~,~~~
\abs{\psi_{l}} =\mathcal{O} \left( {\epsilon \sqrt{\epsilon} \over \rho^3}\right),
\eee
corresponding to the scalings for $u_k$ and $u_l$
\bee
\abs{u_{k}} =\mathcal{O} \left({{\epsilon}\over \rho}\right)~,~~~
\abs{u_{l}} =\mathcal{O} \left({\epsilon^2  \over \rho^3}\right),
\eee
similarly to the scalings for $\omega=0$ (\ref{amp_uk},\ref{amp_ul}).

As above, for simplicity and without loss of generality, we assume an 
excitation of node $j=1$. The evolution of the nodes $\{2,\dots,N\}$ is described by the forced system of linear ordinary differential equations
\begin{equation}
\label{psin}
{\ud \boldsymbol{\varphi} \over \ud t} 
={-\ic \epsilon \over 2\omega} ~\left( \mathbf{\Delta}^{1} ~ \boldsymbol{\varphi}  
+\mathbf{f} \right),
\end{equation}
where $ \boldsymbol{\varphi} =\left(\psi_2,\psi_3,\dots,\psi_N \right)^T$, 
and where $\mathbf{f}=\left(f_1,f_2,\dots,f_{N-1} \right)^T$ is the forcing term such that 
$f_{k}=\psi_1$ if $k$ adjacent to $1$ ($k\sim 1$)
and $0$ otherwise. \\
We expand $\boldsymbol{\varphi}$ using a basis of the eigenvectors 
$\mathbf{z}^{m}$ of $\mathbf{\Delta}^{1}$
\be \label{varphi}
\boldsymbol{\varphi} = \sum_{m=1}^{N-1} \beta_m ~\mathbf{z}^{m}. \ee
Substituting (\ref{varphi}) into (\ref{psin}) and projecting on each eigenvector $\mathbf{z}^{m}$, 
we get
\be
{\ud \beta_m \over \ud t} = \ic {\epsilon \over 2 \omega} \omega_m^2 \beta_m - \ic {\epsilon \over 2 \omega} \psi_1 \sum_{k \sim 1}  z^m_{k-1}.
\ee
The harmonic frequencies of $\boldsymbol{\varphi}$ are $\sqrt{\epsilon \over 2\omega} \omega_m $ for $m\in \{1,\dots,N-1 \}$ such that $\sum_{k \sim 1}  z^m_{k-1} \neq 0$. The harmonic frequencies of $\mathbf{v}=\left(u_{2},u_{3},\dots,u_{N} \right)^T$ using (\ref{psi}) are 
\be
\label{freqnn_nls}
\sqrt{\epsilon \over 2\omega}~ \omega_m + \omega,
\ee
for $m\in \{1,\dots,N-1 \}$ such that $\sum_{k \sim 1}  z^m_{k-1} \neq 0$.
Notice that (\ref{freqnn_nls}) and (\ref{freqnn_phi4}) are almost equal
for large $\omega$ and small $\epsilon$, and these are the conditions 
of validity of the approximation by modulation theory.

\section{Numerical results}
\label{sec4}

\begin{figure}[H]
 \centering
\resizebox{5 cm}{4 cm}{
\includegraphics{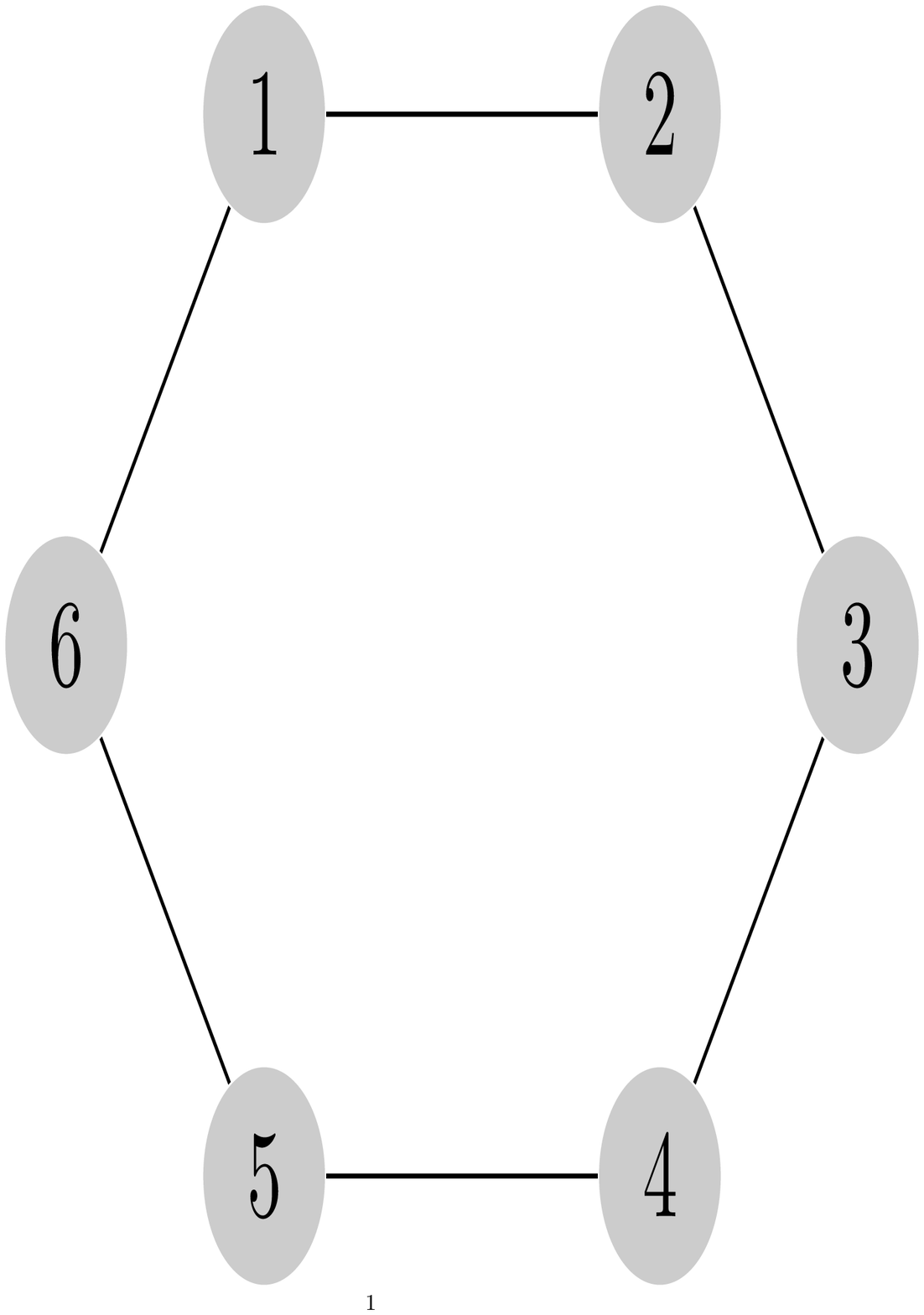}}
\vspace{40pt}
 \caption{\label{cycle6} Cycle 6}
\end{figure}

We illustrate our findings on two graphs: 
a cycle 3 joined to a single isolated node known as the paw graph (Fig.\ref{paw}),
and the cycle 6 (Fig.\ref{cycle6}). The first graph is not regular and
has one symmetry, the permutation of nodes 3 and 4. The second is a cycle 
invariant under cyclic permutations. We will see how symmetries affect
the observed modes and how localized solutions destabilize.

The system of ordinary differential equations (\ref{phi4}) is solved
in double precision, using a Runge-Kutta 4-5 method
with a time step $10^{-2}$ and a relative error of $10^{-8}$.
To check the validity of the solutions, we calculated the
Fourier transform $\widehat{u}_k$ of each $u_k,~ k\in\{1,\dots,N\}$.
This revealed the frequencies of the motion and allowed a detailed comparison 
with the analysis of sections \ref{sec2} and \ref{sec3}.
In practise, we used the fast Fourier transform (FFT) of Matlab
on a time-series of $n=20000$ points on a time $t_f =200$ to approximate the 
continuum Fourier transform. The data was multiplied by a Hamming window
$$u_k (m)\times \left(0.54-0.46 \cos\left(2\pi {(m-1)\over n}\right)\right), $$
for $m\in \{1,\dots,n \} ,~ k\in \{ 1,\dots,N \}.$

\subsection{Paw graph}
\label{subsection_paw}

\begin{figure}[H] 
\centering
\resizebox{4 cm}{2.8 cm}{\includegraphics{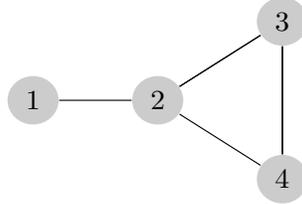}}
\vspace{20pt}
\caption{Paw graph.}
\label{paw}
\end{figure}

We consider the paw graph studied in \cite{ac14}, shown in Fig.\ref{paw}.
This graph is invariant by the permutation symmetry of nodes $3$ and $4$.
Then the components $u_3(t)$ and $u_4(t)$ of the solution $\mathbf{u}(t)$
are equal.

\subsubsection{Natural frequency $\omega=0$}
\subsubsection*{Exciting node 1}

\begin{figure}[H]
\centering 
\resizebox{12 cm}{5 cm}{ 
\includegraphics{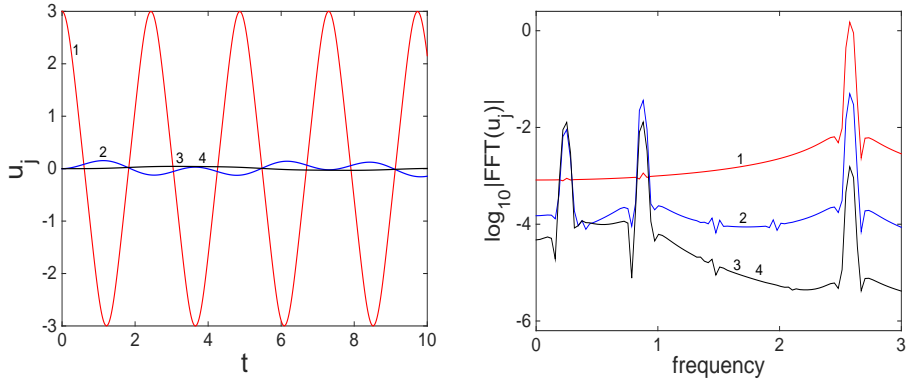}}
\vspace{10pt}
\caption{
Solution of equation (\ref{phi4}) for the paw graph and an initial condition
$u_1(0)=3,~ u_2(0)=u_3(0)=u_4(0)=0$.
Left panel: time evolution of $u_1$ (red online), $u_2$ (blue online), 
$u_3 = u_4$ (black online).
Right panel: Fourier transform of the solutions 
$\widehat{u}_1$ (red online), $\widehat{u}_2$ (blue online), 
$\widehat{u}_3= \widehat{u}_4$ (black online). The 
parameters are  $\omega=0,~\epsilon=0.2$. }
\label{paw_w0} 
\end{figure}

We solve equation (\ref{phi4}) with $\omega=0,~\epsilon=0.2$ for the paw graph. The left panel of Fig.\ref{paw_w0} shows the time evolution of the solutions
$u_k,~k\in\{1,\dots,4\}$ 
when exciting the system at node $j=1$ with initial amplitude $u_1(0)=\rho=3$.
We see clearly a localized solution; it can be observed over more than 
a thousand periods with no significant decay.
The logarithm with base $10$ of the modulus of the 
discrete Fourier transform of the solutions
$\log_{10}( \abs{\widehat{u}_k}) ,~~ k\in\{1,\dots,4\}$ are shown on the 
right panel of Fig.\ref{paw_w0}. The Fourier components of the neighbor
$u_2$ corresponding to the linear mode and the nonlinear mode are
about equal. As we go to the next nearest neighbor the Fourier component
due to the nonlinear excitation of node 1 is a 100 times smaller than
the linear response of the network. This is a general feature that we
see on all the systems we have analyzed. It confirms the exponential 
localization of the nonlinear mode. 

From the Fourier spectrum, we determine that
$u_1$ oscillates at the nonlinear frequency (\ref{Omega0}) $\Omega_0 ={6\pi\sqrt{\pi}  \over \Gamma^2({1\over 4})} \approx 2.54$ 
and at the odd harmonics of $\Omega_0$ ($3 \Omega_0$ and weakly at $5\Omega_0$) due to the Fourier expansion of the solution
(\ref{uj0cn}) (formula (\ref{fcn}) Appendix.\ref{jacobi})
\begin{align*}
u_1(t)&=\rho ~ \mathrm{cn}\left(\rho t,{1\over \sqrt{2}}\right) \\
&\approx
4\sqrt{2} ~\Omega_0
\left[ b_1
\cos\left(\Omega_0 t \right) +\right. +
\left. 
b_3
\cos\left(3\Omega_0 t \right) +
b_5
\cos\left(5\Omega_0 t \right)+\dots \right],
\end{align*}
where 
$$b_1={\mathrm{e}^{-\pi \over 2} 
\over 1+\mathrm{e}^{-\pi} },~~b_3 = {\mathrm{e}^{-3\pi \over 2} 
\over 1+\mathrm{e}^{-3\pi} }, ~~b_5={\mathrm{e}^{-5\pi \over 2} 
\over 1+\mathrm{e}^{-5\pi} }.$$

The solutions shown in Fig.\ref{paw_w0} are such that $u_1 = \mathcal{O}(\rho)$
and $\abs{u_2}=\mathcal{O}\left(\epsilon \over \rho \right),~~ 
\abs{u_3}=\abs{u_4}=\mathcal{O}\left(\epsilon \over \rho^3 \right)$. 
To describe the evolution
of $u_2,u_3$ and $u_4$, it is then natural to reduce the system
(\ref{phi4}) to the linear system forced by $u_1$
\be
\label{u1}
{\ud^2 u_1 \over \ud t^2}=-u_1^3,
\ee
\begin{equation}
\begin{array}{l c r}
\label{u234}
{\ud^2 \over \ud t^2}
\begin{pmatrix}
u_2 \\
u_3 \\
u_4
\end{pmatrix}
= \epsilon
\begin{pmatrix}
-3 & 1 & 1 \\
1 & -2 & 1 \\
1 & 1 & -2 
\end{pmatrix}
\begin{pmatrix}
u_2 \\
u_3 \\
u_4
\end{pmatrix}
+
\begin{pmatrix}
\epsilon u_1 \\
0 \\
0
\end{pmatrix}
\end{array}
\end{equation}

The Fourier representation shows that the nearest neighbor 
$u_2$ and similarly the next nearest neighbors $u_3$ and $u_4$ 
oscillate at the nonlinear 
frequency $\Omega_0$ and at the eigenfrequencies (\ref{freqnn_w0}) of the matrix $\epsilon \mathbf{\Delta}^{1}$
\begin{align*}
\sqrt{\epsilon}\omega_1 &= \sqrt{0.2 \left( 2-\sqrt{3} \right)}\approx 0.23,\\
\sqrt{\epsilon}\omega_3 &=\sqrt{0.2 \left( 2+\sqrt{3} \right)} \approx 0.86,
\end{align*}
where 
\begin{equation}
\label{Dpaw}
\mathbf{\Delta}^{1}=
\begin{pmatrix}
-3 & 1 & 1 \\
1 & -2 & 1 \\
1 & 1 & -2 
\end{pmatrix}.
\end{equation}

The eigenvectors of $\mathbf{\Delta}^{1}$ are
\begin{align}
\mathbf{z}^{1}={1 \over \sqrt{6-2\sqrt{3}}}
\left( 
\begin{array}{c}
\sqrt{3}-1\\
1\\
1\\
\end{array} 
\right),~ \mathbf{z}^{2}
= \frac{1}{\sqrt{2}} \left( 
\begin{array}{c}
0 \\
1 \\
-1 \\
\end{array} 
\right),~
\mathbf{z}^{3}
= {1 \over \sqrt{6+2\sqrt{3}}}
\left( 
\begin{array}{c}
\sqrt{3}+1\\
-1\\
-1\\
\end{array} 
\right)
\label{zpaw}.
\end{align}
The absence of the eigenfrequency $\sqrt{\epsilon} \omega_2=\sqrt{0.2 \times 3} \approx 0.77$
is due to 
$$\sum_{k \sim 1} z_{k-1}^2 =z_1^2 = 0.$$

The equations (\ref{alpha}) are then
\begin{align} 
{\ud^2 a_1 \over \ud t^2}= -0.2\left(2-\sqrt{3} \right) a_1 
+{0.2 (\sqrt{3}-1) \over \sqrt{6-2\sqrt{3}}}  u_1, \label{pawda1}
\\
{\ud^2 a_3 \over \ud t^2}= -0.2\left(2+\sqrt{3} \right) a_3 
+{0.2 (\sqrt{3}+1) \over \sqrt{6+2\sqrt{3}}}  u_1, \label{pawda3}
\end{align}
To conclude, when exciting node 1 with a large amplitude $\rho$, the
evolution of $u_1$ is given by (\ref{uj0cn}) and the evolution
of $u_2, ~u_3$ and $u_4$ by
\be\label{pawdu234}
\begin{pmatrix} u_2(t) \\u_3(t) \\u_4(t) \end{pmatrix} =
a_1 (t) \mathbf{z}^1 + a_3 (t) \mathbf{z}^3 ,\ee
where $a_1$ and $a_3$ are solutions of equations 
(\ref{pawda1},\ref{pawda3}).
{Note that solving the reduced system (\ref{u1},\ref{u234}) yields the same results as the ones shown in Fig.\ref{paw_w0}}

\subsubsection*{Exciting node 2}

\begin{figure}[H]
\centering
\resizebox{12 cm}{5 cm}{
\includegraphics{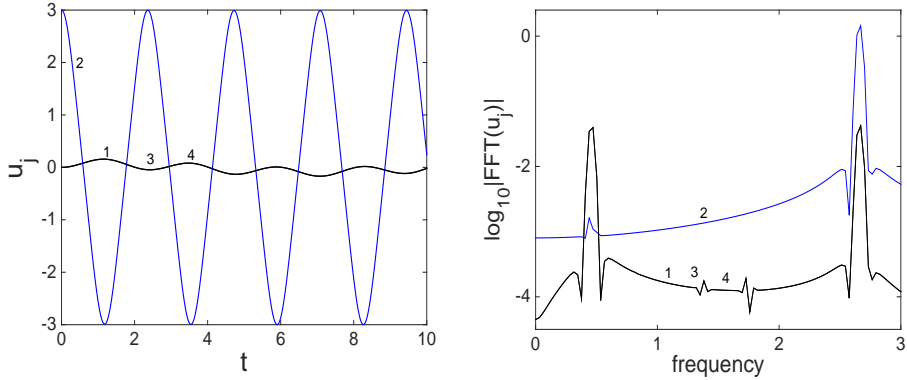}}
\caption{
Solution of equation (\ref{phi4}) for the paw graph and an initial condition
$u_2(0)=3,~ u_1(0)=u_3(0)=u_4(0)=0$.
Left panel: time evolution of $u_2$ (blue online), 
$u_1=u_3=u_4$ (black online).
Right panel: Fourier transform of the solutions 
$\widehat{u}_2$ (blue online), 
$\widehat{u}_1=\widehat{u}_3=\widehat{u}_4$ (black online). The 
parameters are  $\omega=0,~\epsilon=0.2$. }
\label{paw0_node2} 
\end{figure}

To observe the different response of the system, we now
excite node $j=2$ with initial amplitude $u_2(0)=\rho=3$ and 
$\omega=0,~\epsilon=0.2$. 
The left panel of Fig.\ref{paw0_node2} shows the time evolution of the solutions
$u_k,~k\in\{1,\dots,4\}$.
The logarithm with base $10$ of the modulus of the 
discrete Fourier transform of the solutions
$\log_{10}( \abs{\widehat{u}_k}) ,~~ k\in\{1,\dots,4\}$ are shown on the right panel of Fig.\ref{paw0_node2}.

The Fourier spectrum shows that
$u_2$ oscillates at the nonlinear frequency (\ref{Omega0}) $\Omega_0 ={6\pi\sqrt{\pi}  \over \Gamma^2({1\over 4})} \approx 2.54$ 
and at the odd harmonics of $\Omega_0$ ($3 \Omega_0$ and weakly at $5\Omega_0$).
The solutions shown in Fig.\ref{paw0_node2} are such that 
$u_2 = \mathcal{O}(\rho)$
and $\abs{u_k} \ll \rho $ for $k=1,3$ and $4$. To describe the evolution
of $u_1,u_3$ and $u_4$, it is then natural to reduce the system
(\ref{phi4}) to the linear system forced by $u_2$
\begin{equation}
\begin{array}{l c r}
\label{u134}
{\ud^2 \over \ud t^2}
\begin{pmatrix}
u_1 \\
u_3 \\
u_4
\end{pmatrix}
= \epsilon
\begin{pmatrix}
-1 & 0 & 0 \\
0 & -2 & 1 \\
0 & 1 & -2 
\end{pmatrix}
\begin{pmatrix}
u_1 \\
u_3 \\
u_4
\end{pmatrix}
+\epsilon u_2
\begin{pmatrix}
1 \\
1 \\
1
\end{pmatrix}
\end{array}.
\end{equation}

The Fourier representation shows that the nearest neighbors 
$u_1,~u_3$ and $u_4$
oscillate at the nonlinear 
frequency $\Omega_0$ and at the eigenfrequencies (\ref{freqnn_w0}) of the matrix $\epsilon \mathbf{\Delta}^{2}$
\begin{equation}
\sqrt{\epsilon}\omega_1  =
\sqrt{\epsilon}\omega_2 = \sqrt{0.2} \times 1 \approx 0.447,
\end{equation}
where 
\begin{equation}
\mathbf{\Delta}^{2}=
\begin{pmatrix}
-1 & 0 & 0 \\
0 & -2 & 1 \\
0 & 1 & -2 
\end{pmatrix}.
\end{equation}

The eigenvectors of $\mathbf{\Delta}^{2}$ are
\begin{align*}
\mathbf{z}^{1}&={1 \over \sqrt{2}}
\left( 
\begin{array}{c}
0 \\
1\\
1\\
\end{array} 
\right),~~ \mathbf{z}^{2}
=  \left( 
\begin{array}{c}
1 \\
0 \\
0 \\
\end{array} 
\right),~~
\mathbf{z}^{3}
= {1 \over \sqrt{2}}
\left( 
\begin{array}{c}
0 \\
1\\
-1\\
\end{array} 
\right).
\end{align*}

Here, only one linear frequency exists. 
The absence of the eigenfrequency $\sqrt{\epsilon} \omega_3= \sqrt{0.2 \times 3 }$
is due to 
\bee
\sum_{k\sim 2, k<2} z_k^3 + \sum_{k\sim 2, k>2} z_{k-1}^3 =
z^3_1+z^3_2+z^3_3 =0. 
\eee

{Then, the evolution 
of the nearest neighbors $u_1, ~u_3$ and $u_4$ is given by
\be
\begin{pmatrix} u_1(t) \\u_3(t) \\u_4(t) \end{pmatrix} =
a_1 (t) \mathbf{z}^1 + a_2 (t) \mathbf{z}^2 = \begin{pmatrix} a_2(t) \\ {1\over \sqrt{2}} a_1(t) \\{1\over \sqrt{2}} a_1(t) \end{pmatrix} ,\ee
where $a_1$ and $a_2$ are solutions of equations 
\begin{align} 
{\ud^2 a_1 \over \ud t^2}&= -\epsilon \omega_1^2 a_1 
+{2 \over \sqrt{2}} \epsilon u_2, 
\\
{\ud^2 a_2 \over \ud t^2}&= -\epsilon \omega_1^2 a_2 
+ \epsilon u_2.
\end{align}
This explains why the solution $u_1$ is equal to $u_3$ and $u_4$.}

\subsubsection*{Exciting node 3}

\begin{figure}[H]
\centering
\resizebox{12 cm}{5 cm}{
\includegraphics{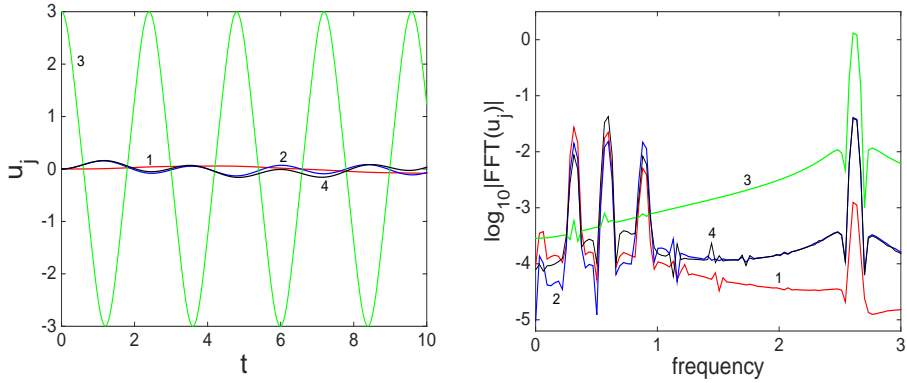}}
\caption{
Solution of equation (\ref{phi4}) for the paw graph and an initial condition
$u_3(0)=3,~ u_1(0)=u_2(0)=u_4(0)=0$.
Left panel: time evolution of $u_1$ (red online), $u_2$ (blue online), 
$u_3$ (green online) and $u_4$ (black online).
Right panel: Fourier transform of the solutions 
$\widehat{u}_1$ (red online), $\widehat{u}_2$ (blue online), 
$\widehat{u}_3$ (green online) and $\widehat{u}_4$ (black online). The 
parameters are  $\omega=0,~\epsilon=0.2$. }\label{paw0_node3} 
\end{figure}

We now excite node 3
with initial amplitude $u_3(0)=\rho=3$ and $\omega=0,~\epsilon=0.2$. 
The left panel of Fig.\ref{paw0_node3} shows the time evolution of the solutions
$u_k,~k\in\{1,\dots,4\}$.
The logarithm with base $10$ of the modulus of the 
discrete Fourier transform of the solutions
$\log_{10}( \abs{\widehat{u}_k}) ,~~ k\in\{1,\dots,4\}$ are shown on the right panel of Fig.\ref{paw0_node3}.

From the Fourier spectrum, we can determine that
$u_3$ oscillates at the nonlinear frequency (\ref{Omega0}) $\Omega_0 ={6\pi\sqrt{\pi}  \over \Gamma^2({1\over 4})} \approx 2.54$ 
and at the odd harmonics of $\Omega_0$ ($3 \Omega_0$ and weakly at $5\Omega_0$).
The solutions shown in Fig.\ref{paw0_node3} are such that $u_3 = \mathcal{O}(\rho)$
and $\abs{u_k} \ll \rho $ for $k=1,2$ and $4$. To describe the evolution
of $u_1,u_2$ and $u_4$, it is then natural to reduce the system
(\ref{phi4}) to the linear system forced by $u_3$
\begin{equation}
\begin{array}{l c r}
\label{u124}
{\ud^2 \over \ud t^2}
\begin{pmatrix}
u_1 \\
u_2 \\
u_4
\end{pmatrix}
= \epsilon
\begin{pmatrix}
-1 & 1 & 0 \\
1 & -3 & 1 \\
0 & 1 & -2 
\end{pmatrix}
\begin{pmatrix}
u_1 \\
u_2 \\
u_4
\end{pmatrix}
+\epsilon u_3
\begin{pmatrix}
0 \\
1 \\
1
\end{pmatrix}
\end{array}
\end{equation}

The Fourier representation shows that the nearest neighbors
$u_2$ and $u_4$ (and similarly the next nearest neighbor $u_1$)
oscillate at the nonlinear 
frequency $\Omega_0$ and at the eigenfrequencies (\ref{freqnn_w0}) of the matrix $\epsilon \mathbf{\Delta}^{3}$
\begin{align*}
\sqrt{\epsilon}\omega_1  &= \sqrt{0.2 \left(2-2\cos\left(2\pi \over 9\right)\right)} \approx  0.3059 , \\
\sqrt{\epsilon}\omega_2  &= \sqrt{0.2 \left(2-2\cos\left(4\pi \over 9\right)\right)} \approx 0.5749 ,\\
\sqrt{\epsilon}\omega_3  &= \sqrt{0.2 \left(2+2\cos\left(\pi \over 9\right)\right)}\approx 0.8808.
\end{align*}
where 
\begin{equation}
\mathbf{\Delta}^{3}=
\begin{pmatrix}
-1 & 1 & 0 \\
1 & -3 & 1 \\
0 & 1 & -2 
\end{pmatrix} .
\end{equation}
The eigenvectors of $\mathbf{\Delta}^{3}$ are
\begin{align*}
\mathbf{z}^{1}&=
\left( 
\begin{array}{c}
0.844 \\
0.449 \\
0.293 \\
\end{array} 
\right),~~ \mathbf{z}^{2}
=  \left( 
\begin{array}{c}
0.449 \\
-0.293 \\
-0.844 \\
\end{array} 
\right),~~
\mathbf{z}^{3}
= 
\left( 
\begin{array}{c}
0.293 \\
-0.844\\
0.449 \\
\end{array} 
\right).
\end{align*}

\subsubsection{Natural frequency $\omega \neq 0$}

\begin{figure}[H]
\centering
\resizebox{12 cm}{5 cm}{
\includegraphics{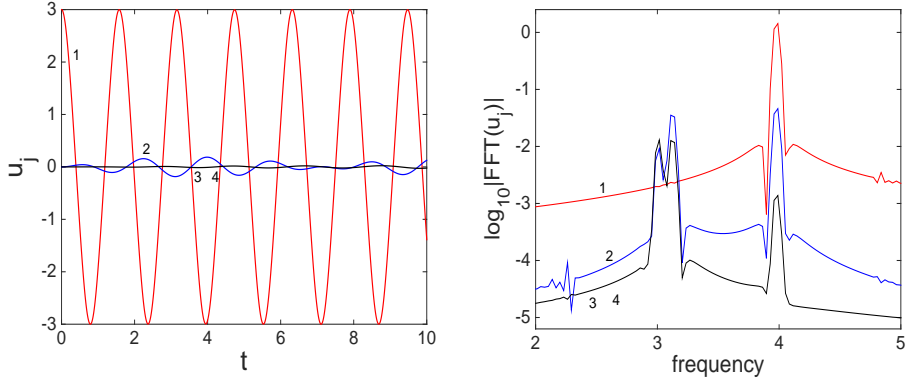}}
       \caption{
Solution of equation (\ref{phi4}) for the paw graph and an initial condition
$u_1(0)=3,~ u_2(0)=u_3(0)=u_4(0)=0$.
Left panel: time evolution of $u_1$ (red online), 
$u_2$ (blue online), $u_3=u_4$ (black online).
Right panel: Fourier transform of the solutions 
$\widehat{u}_1$ (red online), $\widehat{u}_2$ (blue online), 
$\widehat{u}_3=\widehat{u}_4$ (black online).
The parameters are  $\omega=3,~\epsilon=0.2$.}\label{paw_phi4}  
\end{figure}

We now analyze a non zero natural frequency, 
choose $\omega=3$ and $\epsilon=0.2$ and solve 
the graph nonlinear wave equation (\ref{phi4}) 
for the paw graph.
The left of Fig.\ref{paw_phi4} shows the time evolution of 
the localized solutions at $j=1$ with amplitude $u_1(0)=3$. 
Again this has been observed for over a thousand periods
with no significant decay. The right of Fig.\ref{paw_phi4} shows 
the logarithm with base $10$ of the modulus of the discrete Fourier transform of the 
solutions $\log_{10}( |\widehat{u}_k|) ,~~ k\in\{1,\dots,4\}$.

Note that $u_1$ oscillates at frequencies 
$\Omega, ~3 \Omega$ and weakly at $5\Omega$ where $\Omega\approx 4$. 
The nearest neighbor $u_2$ (and similarly the next 
nearest neighbors $u_3$ and $u_4$) oscillate at the nonlinear 
frequency $\Omega$ and at eigenfrequencies (\ref{freqnn_phi4}) 
which are almost equal to those in
(\ref{freqnn_nls})
\begin{align*}
\sqrt{\epsilon \omega_1^2 + \omega^2} &\approx \sqrt{\epsilon \over 2 \omega}\omega_1+ \omega \approx 3, \\
\sqrt{\epsilon \omega_3^2 + \omega^2} &\approx \sqrt{\epsilon \over 2 \omega}\omega_3+ \omega \approx  3.12.
\end{align*}

\begin{figure}[H]
\centering
\resizebox{12 cm}{5 cm}{
\includegraphics{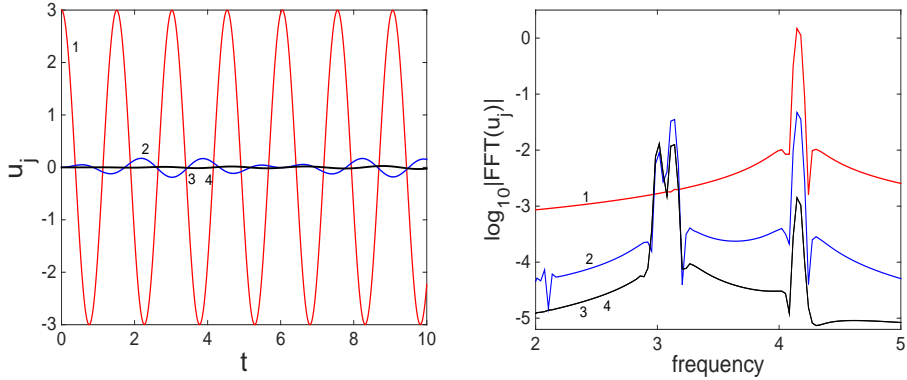}}
       \caption{\label{paw_nls}  
Solution of equation (\ref{gnls}) for the paw graph plotted in the 
$u$ variables using (\ref{psi}) for an initial condition 
$u_1(0)=3,~ u_2(0)=u_3(0)=u_4(0)=0$.
Left panel: time evolution of $u_1$ (red online), $u_2$ (blue online), 
$u_3=u_4$ (black online).
Right panel: Fourier transform of the solutions 
$\widehat{u}_1$ (red online), $\widehat{u}_2$ (blue online), 
$\widehat{u}_3=\widehat{u}_4$ (black online).
Same parameters as in Fig.\ref{paw_phi4}. }
\end{figure}

We proceed to validate the modulation theory
by solving the graph nonlinear Schr\"odinger equation 
(\ref{gnls}) with $\omega=3$ and $\epsilon=0.2$. From this
solution, we calculate $\mathbf{u}$ using the change of variables (\ref{psi}).
The comparison of the left panels of Figs. \ref{paw_phi4} and
\ref{paw_nls} confirm the approximation by the modulation theory. 

When exciting initially node 2, we note that $u_2$ oscillates at frequencies
$\Omega, ~3 \Omega$ and weakly at $5\Omega$ where $\Omega\approx 4$.
The nearest neighbors $u_1,~u_3$ and $u_4$ oscillate at the nonlinear frequency $\Omega$ and at eigenfrequencies (\ref{freqnn_phi4})
\begin{equation*}
\sqrt{\epsilon \omega_1^2 + \omega^2} = \sqrt{\epsilon \omega_2^2 + \omega^2}
\approx 3.033. 
\end{equation*}

Similarly, exciting initially node 3, we observe that $u_3$ oscillates at frequencies
$\Omega, ~3 \Omega$ and weakly at $5\Omega$ where $\Omega\approx 4$.
The nearest neighbors $u_2$ and $u_4$ (and similarly the next nearest neighbor $u_1$)
oscillate at the nonlinear frequency $\Omega$ and at eigenfrequencies (\ref{freqnn_phi4})
\begin{align*}
\sqrt{\epsilon  \omega_1^2 + \omega^2} \approx 3.0156 ,\\
\sqrt{\epsilon  \omega_2^2 + \omega^2} \approx 3.0546 , \\
\sqrt{\epsilon  \omega_3^2 + \omega^2} \approx 3.1266.
\end{align*}

\subsection{Cycle 6}

\subsubsection{Natural frequency $\omega=0$ }
\begin{figure}[H]
\centering
\resizebox{12 cm}{5 cm}{
\includegraphics{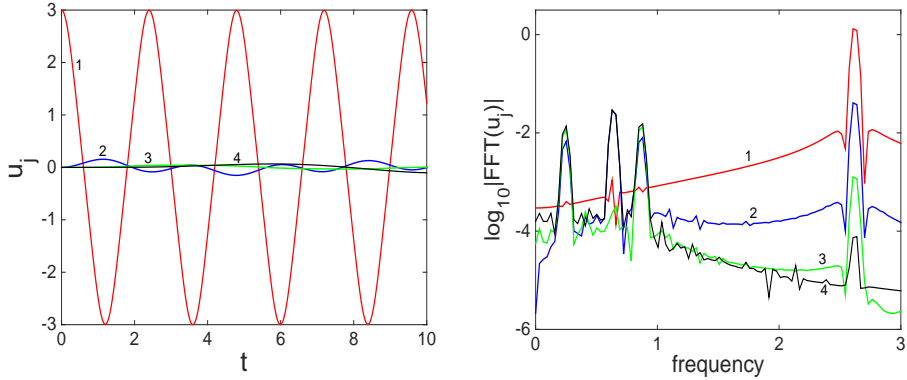}}
       \caption{\label{cy6_w0}  
Left panel: time evolution of $u_1$ (red online), $u_2$ (blue online), $u_3$ (green online) and $u_4$ (black online) solutions of (\ref{phi4}) with $\omega=0,~\epsilon=0.2$ in cycle 6 for initial amplitudes $u_1(0)=3,~ u_2(0)=u_3(0)=u_4(0)=u_5(0)=u_6(0)=0$.
Right panel: Fourier transform of the solutions 
$\widehat{u}_1$ (red online), $\widehat{u}_2$ (blue online), $\widehat{u}_3$ (green online) and $\widehat{u}_4$ (black online).}
\end{figure}

We now consider the cycle 6 where all the nodes are invariant under cyclic permutations, 
so that they behave the same way.
We solve equation (\ref{phi4}) with $\omega=0$ and $\epsilon=0.2$ 
when exciting the system 
at site $j=1$ with initial amplitude $u_1(0)=3$. 
The left panel of Fig.\ref{cy6_w0} shows the time evolution of the solutions
$u_k,~k\in\{1,\dots,6\}$. 
Notice how $u_1$ is large while $u_2,~u_3$ and $u_4$ are small, 
indicating a localized oscillation. 
The right panel of Fig.\ref{cy6_w0}
shows the logarithm with base $10$ of the modulus $\log_{10}( |\widehat{u}_k|) ,~~ k\in\{1,\dots,6\}$ of the 
discrete Fourier transform of the solutions. 
The permutation symmetry of nodes $2 \leftrightarrow 6$ and 
$3 \leftrightarrow 5$ is reflected in the solutions 
$u_2 = u_6$ and $u_3 = u_5$. This network
should show five linear modes, nevertheless due to the symmetry
only three linear modes are present.

The dynamics at nodes $\{2,\dots,6 \}$ is described by
the linear system
\begin{equation}
{\ud^2 \over \ud t^2}
\begin{pmatrix}
u_2 \\
u_3 \\
u_4 \\
u_5 \\ 
u_6
\end{pmatrix}
= \epsilon
\begin{pmatrix}
-2 & 1 & 0 & 0 & 0  \\
1 & -2 & 1 & 0 & 0  \\
0 & 1 & -2 & 1 & 0  \\
0 & 0 & 1 & -2 & 1 \\
0 & 0 & 0 & 1 & -2 \\
\end{pmatrix}
\begin{pmatrix}
u_2 \\
u_3 \\
u_4 \\
u_5 \\ 
u_6
\end{pmatrix}
+
\begin{pmatrix}
\epsilon u_1 \\
0 \\
0 \\
0 \\
\epsilon u_1
\end{pmatrix}
\end{equation}

Using the result of the Fourier spectrum, we can determine that $u_1$ oscillates at frequencies $\Omega_0\approx 2.54$, $3 \Omega_0$ and 
weakly at $5 \Omega_0$. 
The nearest neighbors ($u_2$ and $u_6$) and next nearest neighbors ($u_3$ and $u_5$) oscillate at frequencies $\Omega_0$ and 
at the eigenfrequencies (\ref{freqnn_w0}) of the matrix $\epsilon \mathbf{\Delta}^{1}$

\begin{align*}
\sqrt{\epsilon}\omega_1 &=\sqrt{0.2\left(2-\sqrt{3} \right)} \approx 0.23,\\
\sqrt{\epsilon} \omega_3 &=\sqrt{0.2\times 2} \approx 0.63,\\
\sqrt{\epsilon} \omega_5 &=\sqrt{0.2\left(2+\sqrt{3} \right)} \approx  0.86 ,
\end{align*}
where

\begin{equation}
\label{Dcy6}
\mathbf{\Delta}^{1}=
\begin{pmatrix}
-2 & 1 & 0 & 0 & 0  \\
1 & -2 & 1 & 0 & 0  \\
0 & 1 & -2 & 1 & 0  \\
0 & 0 & 1 & -2 & 1 \\
0 & 0 & 0 & 1 & -2 \\
\end{pmatrix}.
\end{equation}

The eigenvectors of $\mathbf{\Delta}^{1}$ are
\begin{align*}
\mathbf{z}^{1}&=-
{1 \over {2 \sqrt{3}}}
\left( 
\begin{array}{c}
1\\
\sqrt{3}\\
2\\
\sqrt{3} \\
1
\end{array} 
\right),~ \mathbf{z}^{2}
= \frac{1}{2} \left( 
\begin{array}{c}
1 \\
1 \\
0 \\
-1 \\
-1 
\end{array} 
\right),~ \mathbf{z}^{3}
= \frac{1}{\sqrt{3}} \left( 
\begin{array}{c}
1\\
0\\
-1\\
0 \\
1
\end{array} 
\right), \\
\mathbf{z}^{4}
&= \frac{1}{2} \left( 
\begin{array}{c}
-1 \\
1 \\
0 \\
-1 \\
1 
\end{array} 
\right),~~
\mathbf{z}^{5}=
{1 \over {2 \sqrt{3}}}
\left( 
\begin{array}{c}
-1\\
\sqrt{3}\\
-2\\
\sqrt{3} \\
-1
\end{array} 
\right).
\end{align*}

The absence of the frequencies $\sqrt{\epsilon} \omega_2$ and $\sqrt{\epsilon} \omega_4$ is due to 
\begin{align*}
\sum_{l\sim 1} z^2_{l-1} &=z^2_1 + z^2_5 = 0, \\
\sum_{l\sim 1} z^4_{l-1} &=z^4_1 + z^4_5 = 0
\end{align*}

\subsubsection{Natural frequency $\omega \neq 0$}

\begin{figure}[H]
\centering
\resizebox{12 cm}{5 cm}{
\includegraphics{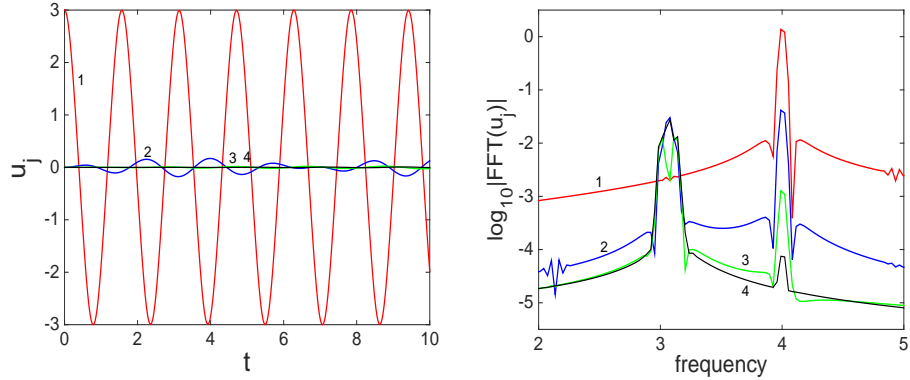}}
       \caption{\label{cy6_phi4}  
Left panel: time evolution of $u_1$ (red online), $u_2$ (blue online), $u_3$ (green online) and $u_4$ (black online) solutions of (\ref{phi4}) with $\omega=3,~\epsilon=0.2$ in cycle 6 for initial amplitudes $u_1(0)=3,~ u_2(0)=u_3(0)=u_4(0)=u_5(0)=u_6(0)=0$.
Right panel: Fourier transform of the solutions 
$\widehat{u}_1$ (red online), $\widehat{u}_2$ (blue online), $\widehat{u}_3$ (green online) and $\widehat{u}_4$ (black online).}
\end{figure}

We consider the equation (\ref{phi4}) with $\omega=3$ and $\epsilon=0.2$ for cycle 6
when exciting the system 
at site $j=1$ with initial amplitude $u_1(0)=3$.
The left panel of Fig.\ref{cy6_phi4} shows the time evolution of 
the localized solutions at $j=1$ with amplitude $u_1(0)=3$. 
Again this has been observed for over a thousand periods
with no significant decay. The right panel of Fig.\ref{cy6_phi4} shows 
the logarithm with base $10$ of the modulus of the discrete Fourier transform of the 
solutions $\log_{10}( |\widehat{u}_k|) ,~~ k\in\{1,\dots,6\}$.

Note that $u_1$ oscillates at frequencies
 $\Omega,~ 3 \Omega$ and weakly at $5\Omega$ where $\Omega\approx 4$.
The nearest neighbors and next nearest neighbors oscillate at frequencies $\Omega$ and at
the eigenfrequencies (\ref{freqnn_phi4}) of the matrix $\epsilon \mathbf{\Delta}^{1} - \omega ^2 \mathbf{I}$
$$\sqrt{\epsilon \omega_1^2 +\omega^2}\approx 3,~~\sqrt{\epsilon \omega_3^2 +\omega^2}\approx 3.06,~~ \sqrt{\epsilon \omega_5^2 +\omega^2}\approx 3.12.$$

Solving the graph nonlinear Schr\"odinger equation 
(\ref{gnls}) with $\omega=3$ and $\epsilon=0.2$ for cycle 6 and calculating $\mathbf{u}$ using the change of variables (\ref{psi}),
yields the same results as the ones shown in Fig.\ref{cy6_phi4}.

\subsection{Weighted graphs}

We now assume that the graph has weights on its edges. 
Consider the paw graph studied in section \ref{subsection_paw} with weights 
$\alpha_{12},~\alpha_{23},~\alpha_{24},~\alpha_{34}$ 
on all the edges 
$\it{e}_{1,2},~\it{e}_{2,3},~\it{e}_{2,4},~ \it{e}_{3,4}$ respectively.
This is done to show the effect of a symmetry break in this graph.
We will see that our formalism extends to this case. 
\begin{figure}[H] 
\centering
\resizebox{4 cm}{2.8 cm}{\includegraphics{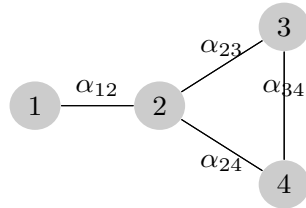}}
\vspace{20pt}
\caption{A weighted paw graph.}
\label{swivelaa}
\end{figure}

The Laplacian $\mathbf{\Delta}$ of the weighted paw graph 
\begin{equation}
\mathbf{\Delta}=
\begin{pmatrix}
-\alpha_{12} & \alpha_{12} & 0 & 0 \\
\alpha_{12} & -\left( \alpha_{12} + \alpha_{23} + \alpha_{24} \right) & \alpha_{23} & \alpha_{24} \\
0 & \alpha_{23} & -\left( \alpha_{23} + \alpha_{34} \right) & \alpha_{34} \\
0 & \alpha_{24} & \alpha_{34} & -\left( \alpha_{24} +   \alpha_{34} \right)
\end{pmatrix}
\end{equation}

The reduced matrix $\mathbf{\Delta}^1$ is
\begin{equation}
\mathbf{\Delta}^1=
\begin{pmatrix}
-\left( \alpha_{12} + \alpha_{23} + \alpha_{24} \right) & \alpha_{23} & \alpha_{24} \\
\alpha_{23} & -\left( \alpha_{23} + \alpha_{34} \right) & \alpha_{34} \\
 \alpha_{24} & \alpha_{34} & -\left( \alpha_{24} +   \alpha_{34} \right)
\end{pmatrix}
\end{equation}

\begin{figure}[H]
\centering 
\resizebox{12 cm}{5 cm}{ 
\includegraphics{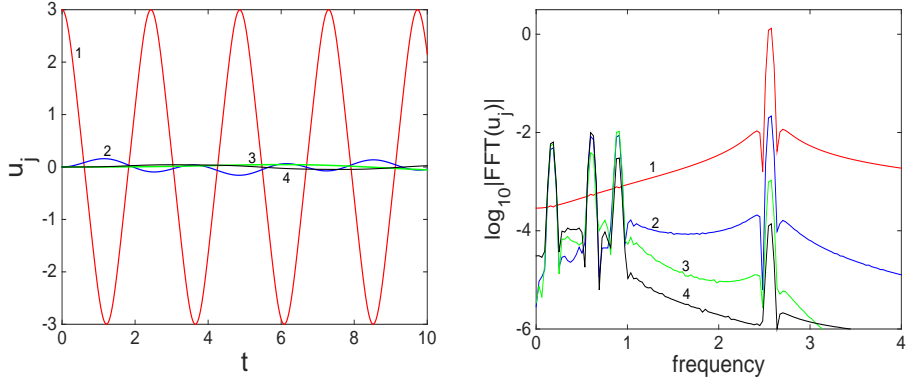}}
\vspace{10pt}
\caption{
Solution of equation (\ref{phi4}) for the paw graph Fig.\ref{swivelaa} with 
$\alpha_{1,2}={1 \over 2},~\alpha_{2,3}={3 \over 2},~\alpha_{2,4}={1 \over 4},~\alpha_{3,4}=1$
and an initial condition
$u_1(0)=3,~ u_2(0)=u_3(0)=u_4(0)=0$.
Left panel: time evolution of $u_1$ (red online), $u_2$ (blue online), 
$u_3$ (green online) and $u_4$ (black online).
Right panel: Fourier transform of the solutions 
$\widehat{u}_1$ (red online), $\widehat{u}_2$ (blue online), 
$\widehat{u}_3$ (green online) and $\widehat{u}_4$ (black online). The 
parameters are  $\omega=0,~\epsilon=0.2$. }
\label{swivelaa1} 
\end{figure}

We solve equation (\ref{phi4}) with $\omega=0,~\epsilon=0.2$ for the paw graph Fig.\ref{swivelaa} with coupling coefficients 
$\alpha_{1,2}={1 \over 2},~\alpha_{2,3}={3 \over 2},~\alpha_{2,4}={1 \over 4},~\alpha_{3,4}=1$.
The left panel of Fig.\ref{swivelaa1} shows the time evolution of the solutions $u_k,~k\in\{1,\dots,4\}$ 
when exciting the system at node $j=1$ with initial amplitude $u_1(0)=\rho=3$.
The right panel of Fig.\ref{swivelaa1}  shows the logarithm with base $10$ 
of the modulus $\log_{10}( |\widehat{u}_k|) ,~~ k\in\{1,\dots,4\}$ of the 
discrete Fourier transform of the solutions. 

Fig.\ref{swivelaa1} shows the exponential decay of the Fourier amplitude 
of the nonlinear component as we go from the excitation node 1
to its nearest neighbor 2 and next nearest neighbor 3, even though the
coupling between 2 and 3 is larger than the
one between 1 and 2. The weight between nodes 2 and 3 is
larger than the one between nodes 2 and 4 so that the symmetry 
between 3 and 4 is broken. Comparing to section \ref{subsection_paw},
here we observe all the eigenfrequencies of $\mathbf{\Delta}^1$
\be
\sqrt{\epsilon} \omega_1 \approx 0.169, ~~~
\sqrt{\epsilon} \omega_2 \approx0.6 , ~~~
\sqrt{\epsilon} \omega_3 \approx0.89.
\ee
The eigenvectors of $\mathbf{\Delta}^1$ are
\begin{align}
\mathbf{z}^{1}=
\left( 
\begin{array}{c}
-0.4939 \\
-0.5867 \\
-0.6417
\end{array} 
\right),~ \mathbf{z}^{2}
= \left( 
\begin{array}{c}
-0.6144 \\
-0.2867  \\
 0.7350   
\end{array} 
\right),~
\mathbf{z}^{3}
= 
\left( 
\begin{array}{c}
 -0.6152 \\
 0.7573  \\
 -0.2189     
\end{array} 
\right)
\label{zpaw2}.
\end{align}

\subsection{Localization vs delocalization}

Up to now, we choose a large amplitude $\rho$
and a small coupling $\epsilon$. This leads to a localized
solution. For a fixed amplitude, as we increase the coupling, the
linear spectrum of the matrix $\mathbf{\Delta}^{j}$ will collide with
the nonlinear frequency $\Omega_0$.
For lattices, this causes the
disappearance of the localized nonlinear solutions, see \cite{MacKay94}
and \cite{flach94}. Aubry and Mackay \cite{MacKay94} point out that localized solutions
also exist for general networks, as long as there is no resonance
with the linear spectrum. This resonance argument explains why the
threshold of the localization depends on the maximal normal eigenfrequency.
Numerically, we observe that the localized solution disappears
and there is a strong coupling with the neighboring nodes.
Also, the spectrum does not show well defined frequencies.

\begin{figure}[H]
\centering
\resizebox{8 cm}{5 cm}
{\includegraphics{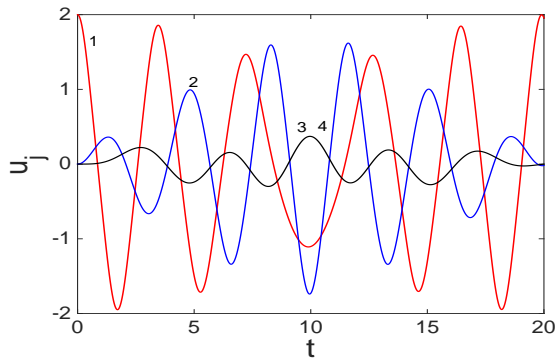}}
\caption{\label{paw_w0eps05} 
Time evolution of $u_1$ (red online), $u_2$ (blue online), $u_3$ and $u_4$ (black online) solutions of (\ref{phi4}) with $\omega=0,~\epsilon=0.5$ in the paw graph for initial amplitudes $u_1(0)=2,~ u_2(0)=u_3(0)=u_4(0)=0$.}
\end{figure}

To illustrate the delocalization regime, we
choose $\omega=0,~\epsilon=0.5$ and solve 
equation (\ref{phi4}) for the paw graph with initial amplitude $u_1(0)=2$.
Fig.\ref{paw_w0eps05} shows the time evolution of the solutions, there
is a strong exchange of energy between nodes.

\begin{figure}[H]
\centering
\resizebox{8 cm}{5 cm}
{\includegraphics{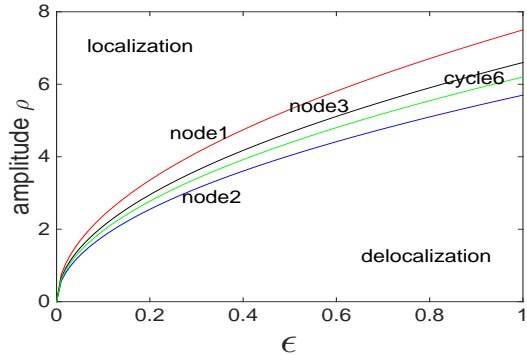}}
\caption{\label{deloc} 
Regions of delocalization in the $\left(\epsilon,\rho \right)$ plane 
with $\omega=0$,
exciting initially nodes 1, 2 and 3 for the paw graph 
and any node for cycle 6
with amplitude $\rho$.
The lines separates the regions of localization (large $\rho$) and the regions of delocalization (small $\rho$).}
\end{figure}

Using the localized character of the solution and the Fourier spectrum
as indicators,
we examined the parameter plane $(\epsilon,\rho)$ and
plotted the regions of localization versus delocalization.
First we consider the paw graph and plot these regions 
for initial excitations of nodes 1, 2 and 3.
This is shown in Fig.\ref{deloc}.

The separation curves behave as $\mathcal{O}(\sqrt{\epsilon})$ since
the eigenfrequencies of $\epsilon \mathbf{\Delta}^{j}$ scale 
like $\sqrt{\epsilon}$.
As one can see, the amplitudes for localization decrease as one
excites node 1 then node 3 and finally node 2. 
The maximal linear frequencies are $\sqrt{\epsilon} \omega_{N-1}$  where
$\omega_{N-1}=1.93$, $1.96$ and $1$, respectively for nodes 1, 3 and 2.
Then the linear spectrum is closer to the nonlinear frequency 
$\Omega_0$ for nodes 1 and 3 and farther for node 2.

We observe that the curve for node 1 is above the one for node 3 
in Fig.\ref{deloc} despite the fact that the maximal linear frequencies 
are very close. This could be due to their different degrees, 
$d_1=1$ and $d_3=2$. Similarly, the curve for node 1 ($d_1=1$) of
the paw graph is above the one for  any node in cycle 6 ($d=2$), 
see Fig.\ref{deloc}, even though the maximal eigenfrequencies are equal.
For equal degrees,  exciting node 3 ( $d_3=2$) and  any node of cycle 6 
gives very close curves in Fig.\ref{deloc}.

Taking $\omega \neq 0$ leads to very close results as 
the linear spectrum and the nonlinear frequency are both approximately
shifted by $\omega$.
When choosing  a quintic nonlinearity in equation (\ref{phi4}),
the nonlinear frequency is much larger while the linear spectrum
is unchanged. Then, we need to increase $\epsilon$
considerably to observe coupling to the linear modes.

\section{Conclusion}
\label{sec5}

We studied localized solutions for a nonlinear graph wave
equation. These are approximated by a nonlinear ordinary 
differential equation for the excited node and a forced 
linear system for the neighboring nodes. 
We validate this approximation by calculating the Fourier spectrum of 
the numerical solution. This shows the nonlinear frequency of the excited node
together with the normal eigenfrequencies of the linear system describing
the neighboring nodes.

The existence of these localized solutions is confirmed using
numerical simulations and 
modulation theory. 
We also examined the localization / delocalization
regions in the parameter plane $(\epsilon, \rho)$. 
Localization holds when the nonlinear frequency $\Omega_0$ is well above 
the linear spectrum which is bounded by 
$\mathcal{O}\left(\sqrt{\epsilon}~ \omega_{N-1} \right)$, where $\omega_{N-1}$ 
is the maximal normal eigenfrequency of $\mathbf{\Delta}^j$.
This condition explains the observed localization amplitude $\rho$.

\section*{Acknowledgment}

This work is part of the XTerM project, co-financed by the European Union 
with the European Regional Development Fund (ERDF) and by the 
Normandie Regional Council. A.B. Aceves thanks the Laboratoire de 
Math\'ematiques de l'INSA de Rouen Normandie for its
hospitality during a visit in 2016. We acknowledge the support
of Agence Nationale de la Recherche through the project Fractal Grid.
  
\begin{appendices}

\section{Basic properties of cosine Jacobi elliptic functions}
\label{jacobi}
Consider the upper limit $\phi$ of the integral
\begin{equation*}
x=\int_{0}^{\phi} {\ud y \over \sqrt{1-\kappa^2 \sin^2 y}}
\end{equation*}
as a function of $x$. 
The function
\begin{equation*}
 \mathrm{cn}\left(x,\kappa\right)=\cos\left(\phi\right), 
\end{equation*}
is called the cosine Jacobi elliptic functions \cite{Abramowitz}, with elliptic modulus $\kappa \in \left[0,1 \right]$.\\
$\mathrm{cn}\left(x,\kappa\right)$ is periodic function with period $4K$, where
\begin{equation*}
K \equiv K(\kappa)= \int_{0}^{\pi \over 2} {\ud \theta \over \sqrt{1-\kappa^2 \sin(\theta)^2}}, 
\end{equation*}
which for $\kappa={1\over \sqrt{2}}$ gives $K\left({1\over \sqrt{2}}\right)= {\Gamma^2\left({1\over 4} \right) \over 4\sqrt{\pi}}\approx 1.8541$.\\
The Fourier series of the cosine elliptic function is given by
\bee
\mathrm{cn}(x,\kappa)={2\pi \over \kappa K} \sum_{m=0}^{\infty} {q^{m+{1\over 2}}\over {1+q^{2m+1}}}
\cos\left((2m+1) {\pi x \over 2K} \right).
\eee
where $q=\mathrm{e}^{-\pi K' \over K}$ so that $q=\mathrm{e}^{-\pi}$ for $\kappa={1\over \sqrt{2}}$.

The Fourier series of the periodic solution ({\ref{uj0cn}}) with period 
$T_0 ={4K\left({1\over \sqrt{2}} \right)\over \rho}={\Gamma^2({1\over 4}) \over \rho \sqrt{\pi}}$
is given by

\begin{align}
u_j(t)=\rho ~ \mathrm{cn}\left(\rho t,{1\over \sqrt{2}}\right) =
{8\pi \sqrt{2} \over T_0} 
\sum_{m=0}^{\infty} b_{2m+1}~
\cos\left((2m+1) {\Omega_0} t \right) \label{fcn},
\end{align}
where $\Omega_0={2\pi \over T_0}$ 
and $$b_{2m+1}={\mathrm{e}^{-\pi \left(m+{1\over 2}\right)} \over {1+\mathrm{e}^{-\pi \left(2m+1\right)}}}.$$

\section{Period of the Duffing oscillator}
\label{appendix_period}
The period of $u_j$ solution of (\ref{uj0}) 
\bee {\ud^2 {u}_j \over \ud t^2}=-u_j^3, \eee
dropping the $j$ index for clarity and integrating the equation, we get
\bee
\left({\ud u \over \ud t}\right)^2 = {1\over 2} \left( \rho^4 - u^4 \right), \eee
where we assumed $u(0)=\rho$ and ${\ud u \over \ud t}(0)=0$.
This yields
\bee  \ud t = \pm \sqrt{2} { \ud u \over \sqrt{\rho^4 - u^4}}. \eee
The period $T_0$ of $u_j(t)$ is then given by the elliptic integral
\bee  
T_0 = 4\sqrt{2} \int_{0}^{\rho} { \ud u \over \sqrt{\rho^4 - u^4}} 
= {4\sqrt{2} \over \rho} \int_{0}^{1} {\ud z \over \sqrt{1-z^4}}.
\eee
We know that
\bee
\int_{0}^{1} {\ud z \over \sqrt{1-z^4}}={\Gamma^2({1\over 4}) \over 4\sqrt{2\pi}}, \eee
where $\Gamma(.)$ is the gamma function and $\Gamma({1\over 4}) \approx 3.6256$.
Then, the period of oscillations is
\bee
T_0 = {\Gamma^2({1\over 4}) \over \rho \sqrt{\pi}} .\eee

\section{Derivation of the Graph nonlinear Schr\"odinger equation}
\label{NLS}
We consider the continuum equation of (\ref{phi4})
\be
\label{nlw}
{\ud ^2 \mathbf{u} \over \ud t^2}+ \omega ^2 \mathbf{u} =\epsilon \mathbf{\Delta} \mathbf{u} - \mathbf{u}^3 .
\ee
To eliminate the term $\omega ^{2} \mathbf{u}$, we write 
$$\mathbf{u}=  \sqrt{\epsilon} \boldsymbol{\psi}(T) \mathrm{e}^{\ic \omega t}
+  \sqrt{\epsilon} \boldsymbol{\psi}^{*}(T) \mathrm{e}^{-\ic \omega t}, $$ 
where $T=\epsilon t$
and $\boldsymbol{\psi}^{*}$ is the complex conjugate of $\boldsymbol{\psi}$.
We have
\begin{align*}
{\ud ^2 \mathbf{u} \over \ud t^2} &=
\epsilon^{5\over 2} \left({\ud  ^2\boldsymbol{\psi}\over \ud T^2}  \mathrm{e}^{\ic \omega t} 
+{\ud ^2 \boldsymbol{\psi}^*\over \ud T^2} \mathrm{e}^{-\ic \omega t}\right)
+ 2\ic \omega \epsilon^{3\over 2} \left( {\ud \boldsymbol{\psi}\over \ud T} \mathrm{e}^{\ic \omega t} 
-{\ud \boldsymbol{\psi}^*\over \ud T} \mathrm{e}^{-\ic \omega t} \right)\\
\\
&-\omega ^2 \sqrt{\epsilon} \left( \boldsymbol{\psi} \mathrm{e}^{\ic \omega t}  
+ \boldsymbol{\psi}^{*} \mathrm{e}^{-\ic \omega t} \right).
\end{align*}
The left hand side of the equation (\ref{nlw}) is
\begin{align*}
{\ud ^2 \mathbf{u} \over \ud t^2}+\omega ^2 \mathbf{u} = 
\epsilon^{5\over 2} \left({\ud  ^2\boldsymbol{\psi}\over \ud T^2}  \mathrm{e}^{\ic \omega t} 
+{\ud ^2 \boldsymbol{\psi}^*\over \ud T^2} \mathrm{e}^{-\ic \omega t}\right)
+ 2\ic \omega \epsilon^{3\over 2} \left( {\ud \boldsymbol{\psi}\over \ud T} \mathrm{e}^{\ic \omega t} 
- {\ud \boldsymbol{\psi}^*\over \ud T} \mathrm{e}^{-\ic \omega t} \right).
\end{align*}
The right hand side of the equation (\ref{nlw}) gives
\begin{align*}
\epsilon \mathbf{\Delta} \mathbf{u} - \mathbf{u}^3 &= 
\epsilon^{3\over 2} \left(\mathbf{\Delta} \boldsymbol{\psi} \mathrm{e}^{\ic \omega t}
+ \mathbf{\Delta} \boldsymbol{\psi}^{*} \mathrm{e}^{-\ic \omega t}
- \boldsymbol{\psi} ^3 \mathrm{e}^{3\ic \omega t} 
-(\boldsymbol{\psi}^{*})^{3}  \mathrm{e}^{-3\ic \omega t} 
\right. \\
\\
&- \left.  3 \abs{\boldsymbol{\psi}}^2 \boldsymbol{\psi}\mathrm{e}^{\ic \omega t}
-3 \abs{\boldsymbol{\psi}}^2 \boldsymbol{\psi}^{*} \mathrm{e}^{-\ic \omega t} \right) .
\end{align*}
We obtain for the order $\mathcal{O}(\epsilon^{3 \over 2})$ 
\begin{align*}
 2\ic \omega  {\ud \boldsymbol{\psi}\over \ud T} \mathrm{e}^{\ic \omega t} 
- 2\ic \omega  {\ud \boldsymbol{\psi}^* \over \ud T} \mathrm{e}^{-\ic \omega t} 
&= \mathbf{\Delta} \boldsymbol{\psi} \mathrm{e}^{\ic \omega t}
+ \mathbf{\Delta} \boldsymbol{\psi}^{*} \mathrm{e}^{-\ic \omega t} 
- \boldsymbol{\psi} ^3 \mathrm{e}^{3\ic \omega t}
- (\boldsymbol{\psi}^{*})^{3} \mathrm{e}^{-3\ic \omega t} \\ 
\\
&
 -3 \abs{\boldsymbol{\psi}}^2 \boldsymbol{\psi}  \mathrm{e}^{\ic \omega t} 
-3  \abs{\boldsymbol{\psi}}^2 \boldsymbol{\psi}^{*}  \mathrm{e}^{-\ic \omega t}.
\end{align*}
Multiplying by $\mathrm{e}^{-\ic \omega t}$, we get
\begin{align*}
2\ic \omega  {\ud \boldsymbol{\psi}\over \ud T}
- 2\ic \omega  {\ud \boldsymbol{\psi}^*\over \ud T} \mathrm{e}^{-2\ic \omega t} &=
\mathbf{\Delta} \boldsymbol{\psi} + \mathbf{\Delta} \boldsymbol{\psi}^{*} \mathrm{e}^{-2\ic \omega t} 
- \boldsymbol{\psi} ^3 \mathrm{e}^{2\ic \omega t} 
-  (\boldsymbol{\psi}^{*})^{3}  \mathrm{e}^{-4\ic \omega t} \\
\\
& - 3 \abs{\boldsymbol{\psi}}^2 \boldsymbol{\psi} 
 -  3 \abs{\boldsymbol{\psi}}^2 \boldsymbol{\psi}^{*}  \mathrm{e}^{-2\ic \omega t}.
\end{align*}
The terms with a non zero phase are rotating fast and average to 
zero on the slow time scale.
Only the terms that have $0$ phase contribute.
This is the {\it rotating wave approximation} \cite{Scott2}.
We obtain the nonlinear Schr\"odinger equation 
\begin{equation*}
2\ic \omega  {\ud \boldsymbol{\psi}\over \ud T} = \mathbf{\Delta} \boldsymbol{\psi} 
- 3 \abs{\boldsymbol{\psi}}^2 \boldsymbol{\psi} .
\end{equation*}

\end{appendices}


\begin{thebibliography}{1}
\bibitem{st88}
A. J. Sievers and S. Takeno, 
\emph{Intrinsic localized modes in anharmonic crystals},
Phys. Rev. Lett. 61, 970-973 (1988).

\bibitem{page}
J. B. Page,
\emph{Asymptotic solutions for localized vibrational modes in strongly anharmonic periodic systems},
Phys. Rev. B 41, 7835-7838 (1990).
       
\bibitem{MacKay94}
R. S. MacKay and S. Aubry,
\emph{Proof of existence of breathers for time-reversible or Hamiltonian networks of weakly coupled oscillators},
Nonlinearity 7, 1623-1643 (1994).
       
\bibitem{flach94}
S. Flach,
\emph{Conditions on the existence of localized excitations in nonlinear discrete systems},
Phys. Rev. E 50, 3134 (1994).
       
\bibitem{flach} 
S. Flach and C. R Willis, 
\emph{Discrete breathers},
Physics reports 295, 181-264 (1998); 
S. Flach and A. V. Gorbach,
\emph{Discrete breathers - Advances in theory and applications},
Physics Reports 467, 1-116 (2008).

\bibitem{Binder}
P. Binder, D. Abraimov, A. V. Ustinov, S. Flach and Y. Zolotaryuk,
\emph{Observation of breathers in Josephson ladders},
Phys. Rev. Lett. 84, 745 (2000).
      
\bibitem{aceves96}
A. B. Aceves, C. De Angelis, T. Peschel, R. Muschall, F. Lederer, S. Trillo, and S. Wabnitz,
\emph{Discrete self-trapping, soliton interactions, and beam steering in nonlinear waveguide arrays}, 
Phys. Rev. E 53, 1172 (1996).
      
\bibitem{eisenberg98}
H. S. Eisenberg, Y. Silberberg, R. Morandotti, A. R. Boyd, and J. S. Aitchison,
\emph{Discrete spatial optical solitons in waveguide arrays},
Phys. Rev. Lett. 81, 3383-3386 (1998). 

\bibitem{fleischer03}
J. W. Fleischer, M. Segev, N. K. Efremidis, and D. N. Christodoulides,
\emph{Observation of two-dimensional discrete solitons in optically induced nonlinear photonic lattices},
Nature (London) 422, 147 (2003).
  
\bibitem{peyrard}
M. Peyrard, \emph{Nonlinear dynamics and statistical physics of DNA},
Nonlinearity 17 R1 (2004).
   
\bibitem{sievers03}
M. Sato, B. E. Hubbard, A. J. Sievers, B. Ilic, D. A. Czaplewski, and H. G. Craighead,
\emph{Observation of locked intrinsic localized vibrational modes in a micromechanical oscillator array},
Phys. Rev. Lett. 90, 044102 (2003).  
     
\bibitem{sievers07}
M. Sato, S. Yasui, M. Kimura, T. Hikihara and A. J. Sievers,
\emph{Management of localized energy in discrete nonlinear transmission lines},
Europhysics Letters 80, 30002 (2007). 
       
\bibitem{christodoulides16}
D. Christodoulides,
\emph{Intrinsic Localized Modes in Optical Photonic Lattices and Arrays},
APS Meeting Abstracts (2016).  

\bibitem{aceves16}
Y. Shen, P. G. Kevrekidis, G. Srinivasan and A. B. Aceves, 
\emph{Existence, stability and dynamics of discrete solitary waves in a binary waveguide array}, 
J. Phys. A: Math. Theor. 49, 295205 (2016).
    
\bibitem{weinstein17} 
M. Jenkinson and M.I. Weinstein,
\emph{Discrete Solitary Waves in Systems with Nonlocal Interactions and the Peierls-Nabarro Barrier},     
Communications in Mathematical Physics Volume 351, Issue 1, 45-94 (2017).

\bibitem{vicencio13}
R.A. Vicencio and M. Johansson,
\emph{Discrete flat-band solitons in the kagome lattice},
Physical Review A 87, 061803 (2013).

\bibitem{kou16}
Y. Kou and J. F\"orstner,
\emph{Discrete plasmonic solitons in graphene-coated nanowire arrays}, 
Optics Express 24, Issue 5, 4714-4721 (2016);
Y. Fan, B. Wang, K. Wang, H. Long and P. Lu. \emph{Plasmonic Zener tunneling in binary graphene sheet arrays},
Optics Letters 41, Issue 13, 2978-2981 (2016).

\bibitem{flach05}
S. Flach and A. Gorbach, 
\emph{Discrete breathers in Fermi-Pasta-Ulam lattices}, 
Chaos 15, 015112 (2005).

\bibitem{panos10}
P. Panayotaros,
\emph{Continuation of normal modes in finite NLS lattices},
Physics Letters A, 374, 3912, (2010).

\bibitem{ovchinnikov}
A. A. Ovchinnikov, N.S. Erikhman, K.A. Pronin, \emph{Vibrational-Rotational
Excitations in Nonlinear Molecular Systems}, Kluwer Academic/Plenum
Publishers, (2001).

\bibitem{Scott1}
A. C. Scott,
\emph{Nonlinear Science: Emergence and Dynamics of Coherent Structures},
Oxford Texts in Applied and Engineering Mathematics, 2nd edn,
Oxford-New York: Oxford University Press (2003).

\bibitem{maas} C. Maas, \emph{Transportation in graphs and the admittance spectrum}, Discrete Applied Mathematics, 16, 32-49, (1987).

\bibitem{cvetkovic}
D. Cvetkovic, P. Rowlinson and S. Simic,
\emph{An Introduction to the Theory of Graph Spectra},
London Mathematical Society Student Texts 75, 
Cambridge: Cambridge University Press (2010).
    
\bibitem{cks13}
J-G. Caputo, A. Knippel and E. Simo, 
\emph{Oscillations of networks: the role of soft nodes},
J. Phys. A: Math. Theor. 46, 035101 (2013).
 
\bibitem{ckkp17}
J-G. Caputo, I. Khames, A. Knippel and P. Panayotaros,
\emph{Periodic orbits in nonlinear wave equations on networks},
J. Phys. A: Math. Theor. 50, 375101 (2017).

\bibitem{Scott2} 
A. C. Scott,
\emph{Encyclopedia of nonlinear science},
London: Routledge, Taylor and Francis Group (2005).

\bibitem{grolet}
A. Grolet, N. Hoffmann, F. Thouverez C. Schwingshackl,
\emph{Travelling and standing envelope solitons in discrete non-linear cyclic structures}.
Mechanical Systems and Signal Processing, Vol 81, 75-87 (2016).

\bibitem{Abramowitz}
M. Abramowitz and I. A. Stegun,
\emph{Handbook of Mathematical Functions}. 
New York: Dover (1965).


\bibitem{Fisk}
Steve Fisk,
\emph{A very short proof of Cauchy's interlace theorem for eigenvalues of Hermitian matrices}.
https://arxiv.org/abs/math/0502408
(2005).

\bibitem{ckk18}
J-G. Caputo, I. Khames and A. Knippel. On graph Laplacians eigenvectors with components in
$\{1,-1,0\}$. "Discrete and Applied Mathematics, in revision, 2018. 
http://arxiv.org/abs/1806.00072

\bibitem{peyrard92}
Y.S. Kivshar and M. Peyrard, 
\emph{Modulational instabilities in discrete lattices},
Phys. Rev. A 46 3198-3205 (1992);
Y.S. Kivshar,
\emph{Localized modes in a chain with nonlinear on-site potential},
Physics Letters A, volume 173, Issue 2, 172-178 (1993). 
     
\bibitem{ac14}
A. B. Aceves and J-G. Caputo, 
\emph{Mode dynamics in nonuniform waveguide arrays: A graph Laplacian approach},
Journal of Optics, Volume 16, Issue 3, article id. 035202 (2014).

\end{thebibliography}
\end{document}